\documentclass[12pt]{article}

\usepackage{xspace,colortbl}
\usepackage{graphicx}
\usepackage{amsmath}
\usepackage{amsthm}
\usepackage{amssymb}
\usepackage{latexsym}
\usepackage{indentfirst}
\usepackage{setspace}
\usepackage[top=1.2in,bottom=1in]{geometry}
\usepackage{mathrsfs}

\begin{document}

\title{\bfseries Exact solution for growth-induced large bending deformation of a hyperelastic plate}

\author{\small Jiong Wang$^1$, Qiongyu Wang$^1$, Hui-Hui Dai$^2$ \footnote{Corresponding author. Email: mahhdai@cityu.edu.hk; Tel.: +852 34428660.}
\\\small $^1$ School of Civil Engineering and Transportation, South China University \\
\small of Technology, 510640 Guangzhou, Guangdong, China
\\\small $^2$ Department of Mathematics, City University of Hong Kong, 83 Tat Chee\\
\small Avenue, Kowloon Tong, Hong Kong}
\date{}
\maketitle

%%%%%%%%%%%%%%%%%%%%%%%%%%%%%%%%%%%%%%%%%%%%%%%%%%%%%%%%%%%%%%%%%%%%%%%%%%%%%%%%%%%%%%%%%%%%%%%%

\begin{abstract}

In this paper, the growth-induced bending deformation of a thin hyperelastic plate is studied. For a plane-strain problem, the governing PDE system is formulated, which is composed of the mechanical equilibrium equations, the constraint equation and the boundary conditions. By adopting a gradient growth field with the growth value changes linearly along the thickness direction, the exact solution to the governing PDE system can be derived. With the obtained solution, some important features of the bending deformation in the plate can be found and the effects of the different growth parameters can be revealed. This exact solution can serve as a benchmark one for testing the correctness of numerical schemes and approximate plate models in growth theory.

\end{abstract}


\section{Introduction}
\label{sec:1}

The growth-induced deformations in soft biological tissues have attracted extensive research interests in the field of biomechanics \cite{hump2002,ben2005,menz2012}. At the continuum level, the mechanical models can be established within the framework of nonlinear elasticity theory \cite{ogden1984}. To describe both the growth effects and the elastic deformations of soft biological materials, the total deformation tensor is usually decomposed into the multiplication of an elastic strain tensor and a growth tensor \cite{rodr1994}. As the elastic deformations of soft biological materials are generally volumetric incompressible, some constraint equation needs to be incorporated in the governing equation system. During the growth processes, mechanical instabilities can be triggered accompanying the large deformations, which result in various pattern formations on the morphogenesis of tissues \cite{li2012}. In the existing works, the instability phenomena in soft materials have been studied through both linear and post-bulking bifurcation analyses \cite{ben2005,audo2008,cao2012}.

Samples with thin plate forms are commonly observed in biological organs or tissues, e.g., leaves, petals, skin. To study the mechanical properties of these samples during the growth process, one usually needs to adopt an appropriate plate theory. In the work of Dervaux and Ben Amar \cite{derv2009}, the classical F$\ddot{o}$ppl-von K$\acute{a}$rm$\acute{a}$n (FvK) plate theory was generalized to study the growth of thin hyperelastic samples. The plate equations were derived through a variational approach, where growth act as a source of mean and Gaussian curvatures. This Fvk-type plate theory has been applied in many different cases and the obtained results can explain some interesting features of the mechanical behaviors of thin hyperelastic samples \cite{derv2010,budd2014,huan2016}. However, applications of the Fvk-type plate theory have some limitations, e.g., only in the range of small strains and some scaling relations need to be satisfied. In the authors' recent work \cite{wangds2016}, a consistent finite-strain plate theory was proposed for incompressible hyperelastic materials, which incorporates both the bending and stretching deformations and satisfies certain consistency requirements with no ad hoc hypotheses. This plate theory will be further developed to study the growth-induced large deformations in thin hyperelastic plates. In our opinion, the different plate theories have their own advantages and should be applied in different situations. To test the correctness of an approximate theory, one sensible thing to do is to make comparisons with some exact solutions. For finite elasticity, some exact solutions are available. However, as far as the authors are aware of, no exact solution for inhomogeneous deformation within the framework of growth theory has been reported. The main aim of this paper is to present the exact solution for growth-induced bending.

More specifically, we shall study the large bending deformation of a hyperelastic plate induced by a gradient growth field. For simplicity, a plane-strain problem is considered and the governing PDE system is formulated, which is composed of the mechanical equilibrium equations, the constraint equation and the boundary conditions. By choosing a specific growth field with the growth value changes linearly along the thickness direction, the exact solution to the governing system can be derived. With the obtained solution, some important features of the bending deformation in the plate can be found and the effects of the different growth parameters can be revealed. We expect that this exact solution can be used as a benchmark problem for evaluating the different plate models for testing the correctness of numerical schemes in growth theory.

This paper is organized as follows. In section 2, the governing PDE system of the model is formulated. In section 3, the exact solution of the governing system is derived. Some further discussions are given in section 4. Finally, some conclusions are drawn.

%%%%%%%%%%%%%%%%%%%%%%%%%%%%%%%%%%%%%%%%%%%%%%%%%%%%%%%%%%%%%%%%%%%%%%%%%%%%%%%%%%%%%%%%%%%%%%%%%%%%%%%%%%%

%%%%%%%%%%%%%%%%%%%%%%%%%%%%%%%%%%%%%%%%%%%%%%%%%%%%%%%%%%%%%%%%%%%%%%%%%%%%%%%%%%%%%%%%%%%%%%%%%%%%%%%%

\section{Governing PDE system}
\label{sec:2}

\begin{figure}
  \begin{minipage}{0.53\textwidth}
  \centering \includegraphics[width=0.95\textwidth]{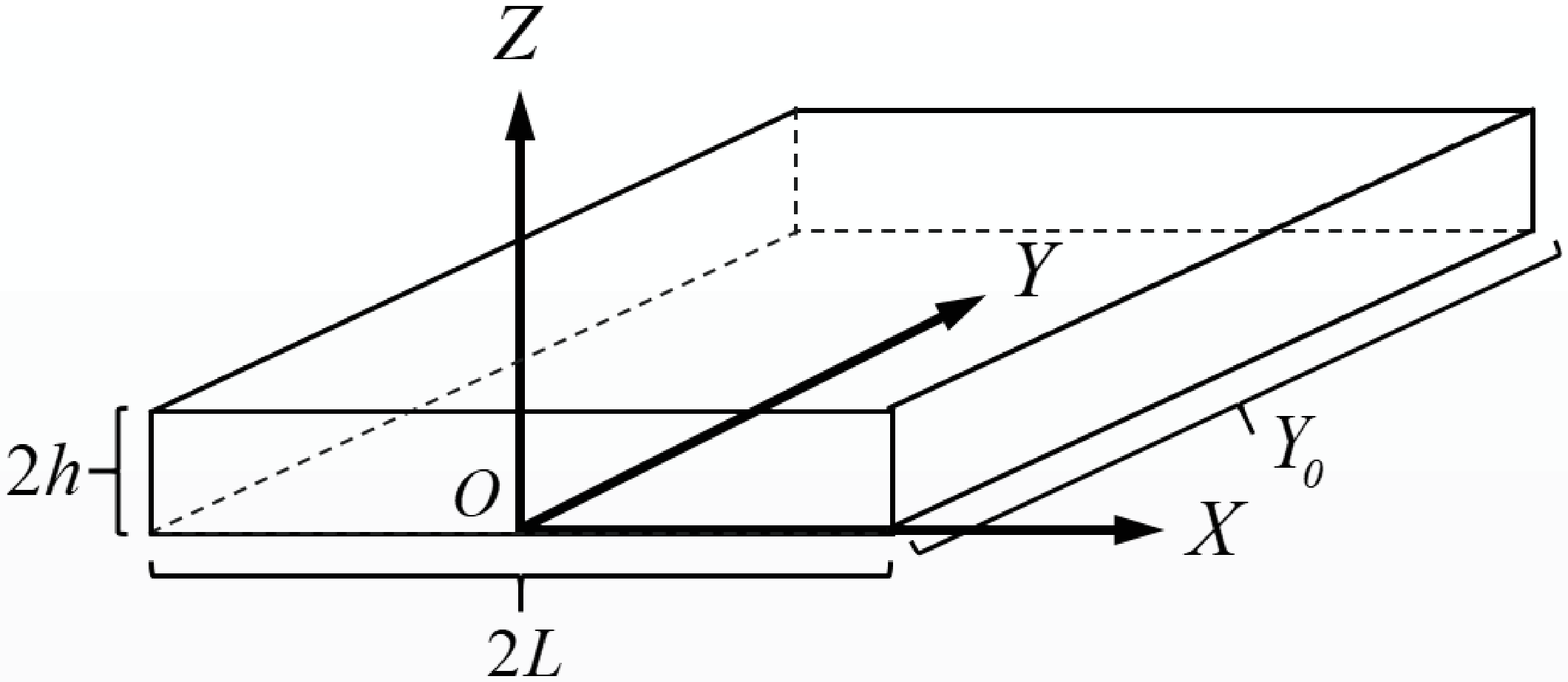}\\(a)\\
  \end{minipage}
  \begin{minipage}{0.45\textwidth}
  \centering \includegraphics[width=0.95\textwidth]{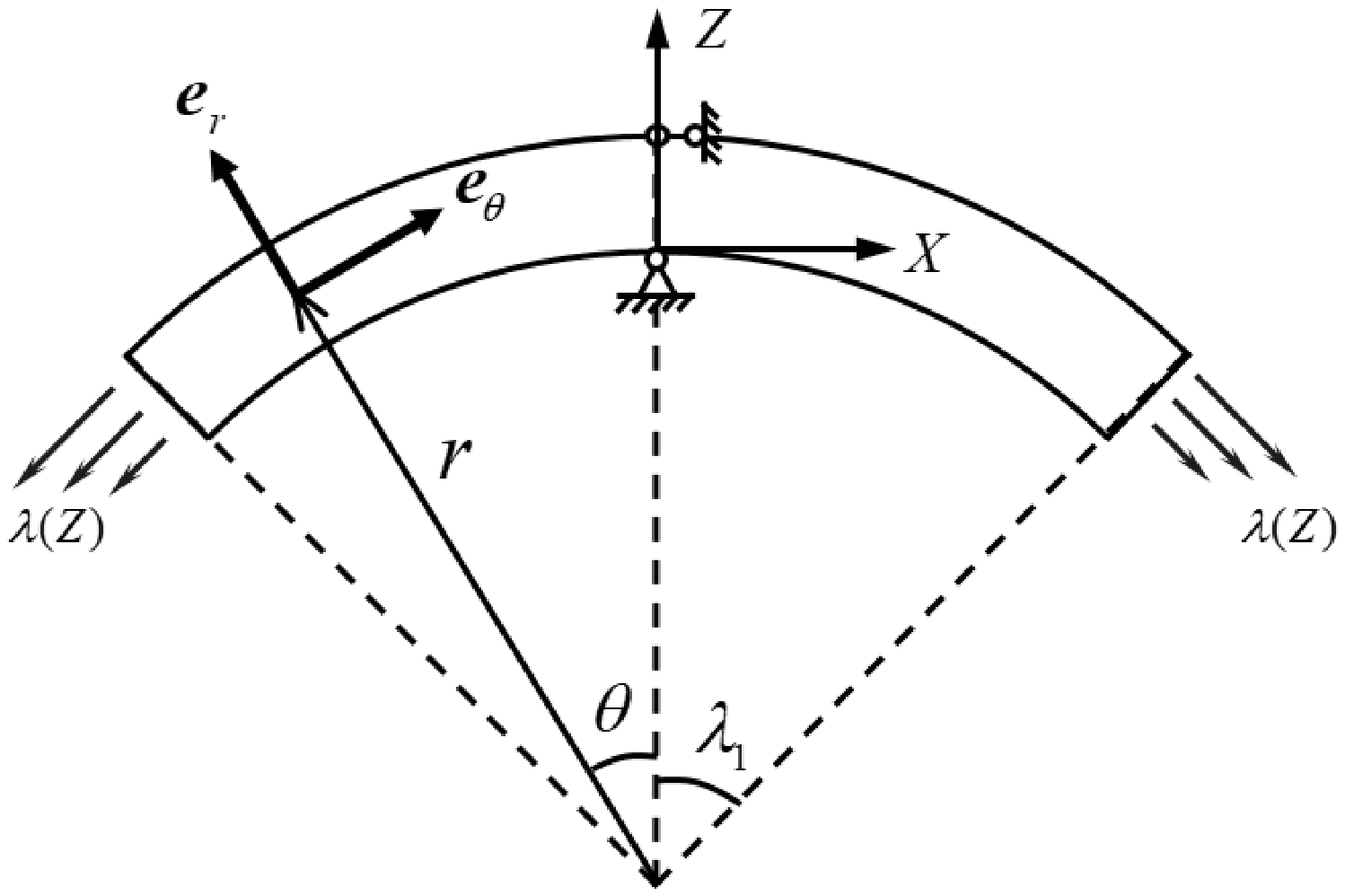}\\(b)\\
  \end{minipage}
  \caption{(a) The 3D reference configuration of the hyperelastic plate; (b) illustration of the bending deformation of the plate induced by the gradient growth field.}
\label{fig:1}
\end{figure}

We consider a thin hyperelastic plate, which occupies the region $\Omega_0=[-L,L]\times[0,Y_0]\times[0,2h]$ in the initial state (cf. Fig. \ref{fig:1}a). The coordinates of a material point in the plate is denoted as $(X,Y,Z)$ in the reference configuration (i.e., the initial state) and $(x,y,z)$ in the current configuration. Suppose that this plate undergoes uniaxial growth along the $X$-axis. The growth tensor can be written as $\mathbb{G}=\mathrm{diag}(\lambda(X,Y,Z),1,1)$, where $\lambda(X,Y,Z)$ is denoted as the growth function. The uniaxial growth also results in the elastic deformation of the plate. Following the approach of Rodriguez et al.\cite{rodr1994}, the total deformation gradient tensor $\mathbb{F}$ can be decomposed into
$$
\mathbb{F}=\mathbb{A}\mathbb{G},\ \ \ \ \eqno(1)
$$
where $\mathbb{A}$ is the elastic deformation tensor. Generally, the elastic deformations of biological soft materials are incompressible, thus we have the constraint equation
$$
R_0(\mathbb{A})=\mathrm{Det}\mathbb{A}-1=0.\ \ \eqno(2)
$$
For simplicity, we suppose that the plate undergoes plane-strain deformations along the $Y$-axis. In this case, we have $y=Y$ and $x$, $z$ and $\lambda$ only depend on the coordinates $X$ and $Z$. The total deformation tensor $\mathbb{F}$ and the elastic deformation tensor $\mathbb{A}$ can then be written as
$$
\mathbb{F}=\left(
\begin{array}{ccc}
x_X & 0 & x_Z \\
0 & 1 & 0 \\
z_X & 0 & z_Z
\end{array}
\right),\ \ \ \
\mathbb{A}=\left(
\begin{array}{ccc}
\frac{x_X}{\lambda} & 0 & x_Z \\
0 & 1 & 0 \\
\frac{z_X}{\lambda} & 0 & z_Z
\end{array}
\right),\ \ \ \
\eqno(3)
$$

To derive some concrete results, we suppose that the plate is made of incompressible neo-Hookean materials, which has the elastic strain-energy function
$$
\phi_0(\mathbb{A})=C_0\left[\mathrm{tr}(\mathbb{A}\mathbb{A}^T)-3\right],\ \ \eqno(4)
$$
where $C_0$ is a material constant. With the expression of $\phi_0$ and by considering the constraint equation (2), the nominal stress tensor can be calculated through \cite{ben2005,derv2009}
$$
\begin{aligned}
\mathbb{S}=&J_G\mathbb{G}^{-1}\left(\frac{\partial\phi_0(\mathbb{A})}{\partial\mathbb{A}}-p\frac{\partial R_0(\mathbb{A})}{\partial\mathbb{A}}\right)=J_G\mathbb{G}^{-1}\left(2C_0\mathbb{A}^T-p\mathbb{A}^{-1}\right)\\
=&\left(\begin{array}{ccc}
-pz_Z+\frac{2C_0x_X}{\lambda} & 0 & px_Z+\frac{2C_0z_X}{\lambda}\\
0 & (2C_0-p)\lambda & 0\\
pz_X+2C_0\lambda x_Z & 0 & -px_X+2C_0\lambda z_Z
\end{array}
\right),
\end{aligned}
\eqno(5)
$$
where $p(X,Z)$ is a Lagrange multiplier and $J_G=\mathrm{Det}\mathbb{G}$ represents the fact that elastic strains are computed from the grown state but not from the reference state.

In the plate region, we have the mechanical equilibrium equation $\mathrm{Div}\mathbb{S}=0$, which yields the following two equations
$$
2C_0x_Z\lambda_Z-p_Xz_Z+p_Zz_X+\frac{2C_0}{\lambda^2}\left(\lambda^3x_{ZZ}-\lambda_Xx_X+\lambda x_{XX}\right)=0,\ \ \eqno(6)
$$
$$
2C_0z_Z\lambda_Z+p_Xx_Z-p_Zx_X+\frac{2C_0}{\lambda^2}\left(\lambda^3z_{ZZ}-\lambda_Xz_X+\lambda z_{XX}\right)=0.\ \ \eqno(7)
$$
With the expression of $\mathbb{A}$ given in (3), the constraint equation (2) can be rewritten as
$$
\frac{1}{\lambda}\left(x_Xz_Z-x_Zz_X\right)-1=0.\ \ \eqno(8)
$$
All the surfaces of the plate are supposed to be traction-free in the $XZ$-plane, thus we have the following boundary conditions
$$
\begin{aligned}
&S_{31}|_{Z=0,2h}=2C_0\lambda x_Z+pz_X|_{Z=0,2h}=0,\\
&S_{33}|_{Z=0,2h}=2C_0\lambda z_Z-px_X|_{Z=0,2h}=0,\\
&S_{11}|_{X=-L,L}=\frac{2C_0x_X}{\lambda}-pz_Z|_{X=-L,L}=0,\\
&S_{13}|_{X=-L,L}=\frac{2C_0z_X}{\lambda}+px_Z|_{X=-L,L}=0.
\end{aligned}
\eqno(9)
$$
Furthermore, to remove the freedom of rigid body motion of the plate, we also need to propose the following restrictions
$$
x(0,0)=z(0,0)=0,\ \ \ \ x(0,2h)=0.\ \ \eqno(10)
$$

Equations (6)-(8) together with (9) and (10) formulate the governing PDE system of the current model. Corresponding to the different growth field distributions $\lambda(X,Z)$, the governing system can be solved to determine the deformations of the plate. In this work, we shall consider the following gradient growth field
$$
\lambda(X,Z)=\lambda_0+\lambda_1Z,\ \ \eqno(11)
$$
and try to derive the exact solution to the governing system.

%%%%%%%%%%%%%%%%%%%%%%%%%%%%%%%%%%%%%%%%%%%%%%%%%%%%%%%%%%%%%%%%%%%%%%%%%%%%%%%%%%%%%%%%%%%%%%%%%%%%%%%%

\section{Derivation of the exact solution}
\label{sec:3}

In this section, we shall derive the exact solution to the governing system (6)-(10) subject to the gradient growth field (11). Without loss of generality, the half length of the plate $L$ is taken to be $1$, then $h$ just represents the thickness-length ratio of the plate. It can be seen that the growth function $\lambda(X,Z)$ given in (11) is independent of $X$ and varies linearly along the thickness of the plate, thus we suppose that the plate will undergo a pure bending deformation. For convenience, we use the cylindrical coordinate system $(\theta,y,r)$ to represent a point in the current configuration (cf. Fig. \ref{fig:1}b). Then, the bending deformation of the plate can be described by
$$
\left\{
\begin{aligned}
&\theta(X,Z)=\alpha X,\ \ \ \ &-L\leqslant X\leqslant L,\\
&r(X,Z)=r(Z),\ \ \ \ &0\leqslant Z\leqslant 2h,\\
&p(X,Z)=p(Z),\ \ \ \ &0\leqslant Z\leqslant 2h,\\
&y=Y, \ \ \ \ &0\leqslant Y\leqslant Y_0,
\end{aligned}
\right.
\eqno(12)
$$
where $\alpha$ is the bending angle to be determined. By using (12), the Cartesian coordinates $\{x,z\}$ can be expressed as
$$
x(X,Z)=r(Z)\sin(\alpha X),\ \ \ z(X,Z)=r(Z)\cos(\alpha X)-r(0).\ \ \eqno(13)
$$
It can be seen that the restrictions given in (10) are automatically satisfied by (13). Substituting (13) into the governing system (6)-(9) and through some manipulations, we obtain the following two equations
$$
-\frac{2C_0\alpha^2r(Z)}{\lambda(Z)}-\alpha r(Z)p'(Z)+2C_0r'(Z)\lambda'(Z)+2C_0\lambda(Z)r''(Z)=0,\ \ \eqno(14)
$$
$$
\frac{\alpha r(Z)r'(Z)}{\lambda(Z)}-1=0,\ \ \eqno(15)
$$
and the boundary conditions
$$
\begin{aligned}
&\lambda(Z)\left(2C_0r'(Z)-\frac{p(Z)}{r'(Z)}\right)|_{Z=0,2h}=0,\ \\
&\frac{2C_0\alpha r(Z)}{\lambda(Z)}-\frac{p(Z)\lambda(Z)}{\alpha r(Z)}|_{X=-1,1}=0.\ \
\end{aligned}
\eqno(16)
$$
Then, the original PDE system (6)-(10) has been reduced to the ODE system (14)-(16).

\noindent\textbf{Remark:} If we use the Cartesian coordinate $(X,Y,Z)$ for a reference point and the cylindrical coordinate system $(\theta,y,r)$ for a current point, the deformation tensors and the nominal stress tensor can be rewritten as
$$
\mathbb{F}=\left(
\begin{array}{ccc}
\alpha r(Z) & 0 & 0 \\
0 & 1 & 0 \\
0 & 0 & r'(Z)
\end{array}
\right),\ \ \ \
\mathbb{A}=\left(
\begin{array}{ccc}
\frac{\alpha r(Z)}{\lambda(Z)} & 0 & 0 \\
0 & 1 & 0 \\
0 & 0 & r'(Z)
\end{array}
\right),\ \ \ \
\eqno(17)
$$
and
$$
\begin{aligned}
\mathbb{S}=&\left(\begin{array}{ccc}
-\frac{2C_0\alpha r(Z)}{\lambda(Z)}-\frac{p(Z)\lambda(Z)}{\alpha r(Z)} & 0 & 0\\
0 & (2C_0-p(Z))\lambda(Z) & 0\\
0 & 0 & \lambda(Z)\left(2C_0r'(Z)-\frac{p(Z)}{r'(Z)}\right)
\end{array}
\right).
\end{aligned}
\eqno(18)
$$
With the expressions of $\mathbb{F}$, $\mathbb{A}$ and $\mathbb{S}$, the mechanical equilibrium equations, the constraint equation and the boundary conditions can be established directly, which are just equivalent to the ODE system (14)-(16).

The ODE system (14)-(16) can be solved directly. First, from $(16)_2$, it can be obtained that
$$
p(Z)=\frac{2C_0\alpha^2r^2(Z)}{\lambda^2(Z)}. \ \eqno(19)
$$
By substituting the expression of $\lambda(Z)$ (i.e. Eq. (11)) into (15), we obtain an ODE of $r(z)$, which can be solved to give
$$
r(Z)=\sqrt{\frac{2\lambda_0}{\alpha}Z+\frac{\lambda_1}{\alpha}Z^2+2c_1},\ \ \eqno(20)
$$
where $c_1$ is a constant to be determined. Further substituting (19) and (20) into $(16)_1$, we obtain two algebraic equations for $\alpha$ and $c_1$, which have the solution
$$
\alpha=\lambda_1,\ \ \ \ c_1=\frac{\lambda_0^2}{2\lambda_1^2}.\ \ \eqno(21)
$$
With the solution given in (19)-(21), it can be checked that equation (14) is automatically satisfied. Thus, we have derived the following solutions to the ODE system (14)-(16)
$$
r(Z)=\frac{\lambda_0}{\lambda_1}+Z,\ \ \ \alpha=\lambda_1,\ \ \ p(Z)=2C_0.\ \ \eqno(22)
$$
By substituting (22) into (13), the exaction solution in the rectangular coordinates can be obtained.

%%%%%%%%%%%%%%%%%%%%%%%%%%%%%%%%%%%%%%%%%%%%%%%%%%%%%%%%%%%%%%%%%%%%%%%%%%%%%%%%%%%%%%%%%%%%%%%%%%%%%%%%

\section{Discussions}
\label{sec:4}

Based on the exact solution obtained in section \ref{sec:3}, we can give some further discussions on the large bending deformation of the plate induced by the gradient growth field (11). First, through some simple analyses, the following features of the bending deformation can be obtained:

\begin{itemize}

\item After the deformation, the configuration of the plate becomes a segment of a circular ring with inner radius $\lambda_0/\lambda_1$ and outer radius $\lambda_0/\lambda_1+2h$, which implies that the thickness of the plate always equals $2h$.

\item The section plane perpendicular to the axial line of the plate (i.e., the $X$-axis) in the initial configuration keeps flat and perpendicular to the axial line after the bending deformation.

\item The layers in the plate with different values of $Z$, including the lower ($Z=0$) and upper ($Z=2h$) surfaces of the plate, have the same bending angle $\alpha=\lambda_1$.

\item During the growth process, no elastic deformation takes place in the plate and the components of the nominal stress tensor are always equal to zero, i.e., no residual stress appears in the plate.

\end{itemize}

\begin{figure}
  \centering \includegraphics[width=0.7\textwidth]{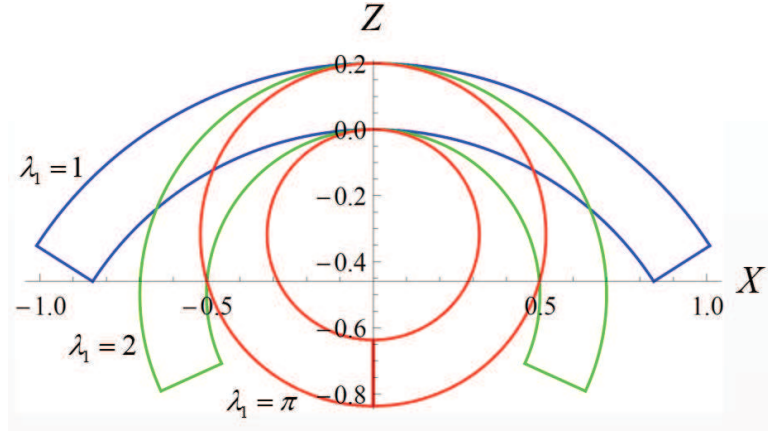}\\(a)\\\vspace{0.3cm}
  \centering \includegraphics[width=0.6\textwidth]{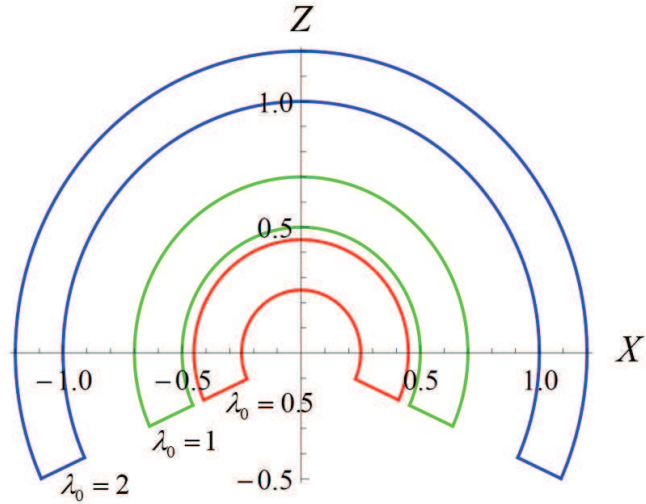}\\(b)\\
  \caption{Configurations of the plate (with thickness value $h=0.1$) corresponding to the different parameter values: (a) $\lambda_0=1$ and $\lambda_1=1$, $2$ and $\pi$; (b) $\lambda_1=2$ and $\lambda_0=0.5$, $1$ and $2$.}
\label{fig:2}
\end{figure}

Besides that, the influences of the growth parameters $\lambda_0$ and $\lambda_1$ on the deformations of the plate can be revealed. In Fig. \ref{fig:2}, the current configurations of the plate corresponding to the different values of $\lambda_0$ and $\lambda_1$ are plotted, where the thickness-length ratio $h=0.1$. From Fig. \ref{fig:2}a, it can be seen that corresponding to the different values of $\lambda_1$, the plate becomes some circular segments with different radii and bending angles. In fact, the bending angle of the plate just equals $\lambda_1$ (cf. Eq. $(21)_1$). The plate keeps flat when $\lambda_1=0$ and becomes a closed circular ring when $\lambda_1=\pi$. From Fig. \ref{fig:2}b, we find that with given value of $\lambda_1$, $\lambda_0$ just represents the elongation ($\lambda_0\geq1$) or shortening ($\lambda_0<1$) of the plate along the axial direction. As mentioned before, the values of $\lambda_0$ and $\lambda_1$ have no influence on the thickness of the circular segment.

%%%%%%%%%%%%%%%%%%%%%%%%%%%%%%%%%%%%%%%%%%%%%%%%%%%%%%%%%%%%%%%%%%%%%%%%%%%%%%%%%%%%%%%%%%%%%%%%%%%%%%%%

\section{Conclusions}
\label{sec:5}

In this work, we studied the growth-induced large bending deformation in a thin plate made of incompressible neo-Hookean materials. For a gradient growth field with the growth value changes linearly along the thickness direction, we derived the exact solution to the governing PDE system. Based on the exact solution, some important features of the bending deformation can be found. The influences of the growth parameters on the current configurations of the plate can also be revealed.

Although the model studied in the current work is simple, it can be viewed as the prototype of many biological samples undergoing natural growth. The results obtained here would be useful for future studies of growth-induced deformations in thin hyperelastic plates. Also, this exact solution can be used as a benchmark problem for validating different approximate plate models and numerical schemes in growth theory.

%%%%%%%%%%%%%%%%%%%%%%%%%%%%%%%%%%%%%%%%%%%%%%%%%%%%%%%%%%%%%%%%%%%%%%%%%%%%%%%%%%%%%%%%%%%%%%%%%%%%%%%%%%%

%%%%%%%%%%%%%%%%%%%%%%%%%%%%%%%%%%%%%%%%%%%%%%%%%%%%%%%%%%%%%%%%%%%%%%%%%%%%%%%%%%%%%%%%%%%%%%%%%%%%%%%%%%%%%%


\begin{thebibliography}{99}

\bibitem{hump2002} Humphrey, J. D., 2002. Continuum biomechanics of soft biological tissues. Proc. R. Soc. Lond. A 459, 3-46.

\bibitem{ben2005} Ben Amar, M., Goriely, A., Growth and instability in elastic tissues. J. Mech. Phys. Solids 53, 2284-2319.

\bibitem{menz2012} Menzel, A., Kuhl, E., 2012. Frontiers in growth and remodeling. Mech. Res. Commun. 42, 1-14.

\bibitem{ogden1984} Ogden, R. W., 1984. Non-linear Elastic Deformation. Dover, Newyork.

\bibitem{rodr1994} Rodriguez, A. K., Hoger, A., McCulloch, A., 1994. Stress-dependent finite growth in soft elastic tissue. J. Biomech. 27, 455-467.

\bibitem{li2012} Li, B., Cao, Y.P., Feng, X.Q., Gao, H.J., 2012. Mechanics of morphological instabilities and surface wrinkling in soft materials: a review. Soft Matter 8, 5728-5745.

\bibitem{audo2008} Audoly, B., Boudaoud, A., 2008. Buckling of a stiff film bound to a compliant substrate: part I-III. J. Mech. Phys. Solids 56, 2401-2458.

\bibitem{cao2012} Cao, Y.P., Hutchinson J. W., From wrinkles to creases in elastomers: the instability and imperfection-sensitivity of wrinkling, Proc. R. Soc. A 468, 94-115.

\bibitem{derv2009} Dervaux, J., Ciarletta, P., Ben Amar, M., 2009. Morphogenesis of thin hyperelastic plates: a constitutive theory of biological growth in the F$\ddot{o}$ppl-von K$\acute{a}$rm$\acute{a}$n limit. J. Mech. Phys. Solids 57, 458-471.

\bibitem{derv2010} Dervaux, J., Ben Amar, M., 2010. Localized growth of layered tissues. IMA J. Appl. Math. 75, 571-580.

\bibitem{budd2014} Budday, S., Steinmann, P., Kuhl, E., 2014. The role of mechanics during brain development. J. Mech. Phys. Solids 72, 75-92.

\bibitem{huan2016} Huang, X., Li, B., Hong, W., Cao, Y. P., Feng, X. Q., 2016. Effects of tension¨Ccompression asymmetry on the surface wrinkling of film¨Csubstrate systems. J. Mech. Phys. Solids 94, 88-104.

\bibitem{wangds2016} Wang, J., Dai, H.-H., Song, Z.L., 2016. On a consistent finite-strain plate theory for incompressible hyperelastic materials. Inter. J. Solids Struct. 78-79, 101-109.

\end{thebibliography}
\end{document}